\documentclass[preprint,12pt]{elsarticle}
\usepackage{amssymb}
\usepackage{graphicx}
\usepackage{natbib} 

\journal{Physics of the Dark Universe}

\begin{document}

\begin{frontmatter}

\title{Ultrafast modulations in stellar, solar and galactic spectra: Dark Matter and numerical ghosts, stellar flares and SETI.}

\author[1]{Fabrizio Tamburini}
\author[2,3,4]{Ignazio Licata}

\address[1]{Rotonium, Le Village by CA, Pz. G. Zanellato, 23, 35131 Padova PD, Italy.}

\address[2]{Institute for Scientific Methodology (ISEM), Palermo, Italy}
            
\address[3]{School of Advanced International Studies on Theoretical and Nonlinear Methodologies of Physics,
 Bari, 70124, Italy}
             
\address[4]{International Institute for Applicable Mathematics and Information Sciences (IIAMIS), B.M. Birla Science Centre, Adarsh Nagar, 500463, Hyderabad, India}

\begin{abstract} 
From new results presented in the literature we discuss the hypothesis that the ultrafast periodic spectral modulations at $f_S \simeq 0.607$ THz found in the spectra of $236$ stars of the Sloan Digital Sky Survey (SDSS) \cite{borra} were due to oscillations induced by dark matter (DM) cores in their centers \cite{tamlicDM} behaving as oscillating boson stars \cite{brito,brito2}.
Two additional frequencies in the redshift-corrected SDSS galactic spectra were found  \cite{borragal}, $ f_{1,G} \simeq 9.5$ THz, the beating between $f_S$ and a spurious frequency, $f_{2,G} \simeq 8.9$ THz, introduced during the data analysis \cite{hippke}. 
The indication that $f_S$ can be real is its detection in a real solar spectrum but not in the Kurucz's artificial solar spectrum \cite{hippke,ku1,ku2}. 
Then, independent SETI observations of four of these stars could not confirm with high confidence, but not completely exclude, the presence of $f_S$ in their power spectra \cite{isaacson} while the radio SETI deep-learning analysis with artificial intelligence confirmed indirectly $f_S$ detecting a narrowband Doppler drifting of radio signals in two of these stars over a sample of $7$ with a high S/N \cite{ma}. Numerical simulations suggest that the drifting can be due to frequency and phase modulation in time of the observed frequencies at $1.3-1.7$ GHz with $f_S$.
This would imply a DM upper mass limit $m_a \lesssim 2.4 \times 10^{3}~ \mathrm{\mu eV}$ \cite{tamlicDM} which also agrees with the results from the gamma ray burst GRB221009A \cite{GRB1,GRB2,GRB3}, laser interferometry \cite{aiello}, suggesting new physics for the muon g-2 anomaly \cite{muon1}.
\end {abstract}



\end{frontmatter}

\section{Introduction}

Dark Matter and Dark Energy (DE) are among the hottest topics in astroparticle physics and modern cosmology \cite{kolbe}. 
While DE is thought to be the main responsible for the accelerated expansion of the Universe, contributing with the $\sim 68\%$ of the total energy \cite{miaoli,clarkson}, DM is supposed to be gravitationally dominating the Universe, accounting for $\sim 85\%$ of the total matter. From Concordance Cosmology DM is the $26.8\%$ of the total mass-energy of the Universe, whilst the ordinary matter is only $4.9\%$ \cite{planck}.
The presence of DM is usually revealed through gravitational lensing, like in the Bullet Cluster, formed after the collision of two clusters of galaxies \cite{clowe,mark}, and explains the discrepancies observed between the separated signatures of gravitational and luminous matter observed in the rotation of galaxies and plays a key role in the evolution of galactic structures and in the formation of clusters of galaxies \cite{zwicky} up to the large-scale structures of the Universe, including the anisotropies in the cosmic microwave background \cite{dm}.

The Big Bang nucleosynthesis \cite{frenk,field}, mainly based on the Standard Model of Particles and supersymmetric (SUSY) theories suggests that DM can be mostly made up of exotic baryonic and non-baryonic matter, with fields that can be either scalar of vectorial. 
An example is the particle axion \cite{ax1,ax2,ax3}, introduced by Peccei and Quinn in 1977 \cite{peccei} to solve the strong CP (charge--parity) problem in quantum chromodynamics.
Other possible candidates are the generalization of the axion, the axion-like particles (ALPS), WIMPs (weakly interacting massive particles), sterile or massive neutrinos and other hypothetical particles that weakly interact with the electromagnetic field \cite{kolbe,duffy,wil,sikvie,dm2,bertone,lux,panda}. 

Axions are pseudo-Nambu-Goldstone bosons with mass $m_a$, a free parameter of the Peccei--Quinn theory, defined in a wide range of masses that depend on the coupling constant of the field, $10^{-6}~\mathrm{\mu eV/c^2}< m_a<1~\mathrm{MeV/c^2}$. 
Recent studies based on lattice QCD simulations that describe the formation of axions during the post-inflationary period within the framework of the standard model of particles, suggest a mass range $5 < m_a < 1500 ~\mathrm{\mu eV/c}^2$; more recent results prefer a mass $m_a \simeq 65~\mathrm{\mu eV/c^2}$ \cite{qcd,qcd2,qcd3} and a recent gamma ray burst explosion GRB221009A indicates a mass on the order of $0.01 - 0.1~\mathrm{\mu eV/c^2}$ \cite{GRB1,GRB2,GRB3}. 
Experimental results from CAST indicates an axion-photon coupling strength $0.66 \times 10^{-10}~\mathrm{GeV}^{-1}$ at $95\%$ confidence level (c.l.) when $m_a \leq 0.02$ eV \cite{cast}. 
ALPs are axion-like particles are an extension of the axion that are present in many extensions of the Standard-Model and in string theory but they are not always introduced in a theoretical framework to solve the strong CP-problem.

Axions and ALPs are supposed to form halos that in most cases surround the visible part of the observed cosmic structures and are expected to be present in our galactic halo \cite{eroncel,chat}. Those particles can be gravitationally captured e.g. by a star and scatter with the nuclei of the elements present there inside, loosing their kinetic energy, with the result of remaining gravitationally trapped in the stellar core. In this scenario, the axionic (or ALP) field is usually thought to behave as a classical field, stable and coherent with respect to gravity and, when trapped inside a star, it can remain gravitationally confined \cite{davidson} forming DM cores with an evaporation rate that depends on the mass of the particle $m_a$ \cite{barranco}. Theory suggests that ALPs ultralight bosonic fields piled up in stars can form temporary stable structures, boson cores, that oscillate in time with a so-called ``breathing'' behavior, typical of exotic objects known as boson stars, where both spacetime and the bosonic fields there present have a stress-energy tensor with characteristic oscillatory modes. Thus, these structures are expected to oscillate at one or more frequencies that depend on the spectrum of masses of the boson field \cite{brito}, with the result of making the host star oscillating \cite{tamlicDM}. Of course, this does not mean that all stars hosting a certain quantity of DM would exhibit a stable observable oscillating behavior in time. 

There are many reasons to think that there oscillatory modes induced by DM should be transient. An example is when the trapped field has a multiple-valued mass spectrum or different particles corresponding to different DM fields are gravitationally trapped in the star. In both cases the oscillations induced by the DM fields are expected to set up a spectrum of oscillatory modes that either stabilize the stellar structure in a static or stable oscillatory dynamical configuration in time in the simplest case, or, because of beatings between different frequencies, these modes generate modulations that can make the star evolving from a stable to a transient oscillatory motion and vice versa or to a route to chaos \cite{brito2}, with the result that any oscillatory motion would be observable only for a finite interval of time.

Another important issue to take in account for the durability of the oscillatory regime of the star is the loss of ALPs from the stellar core. The first cause is simply that cores made with ALPs having a very small mass are expected to evaporate and leave the stellar structure,  the DM core evaporation occurs when the star cannot compensate the process of DM evaporation with the gravitational capture and condensation. The second one is when ALPs are converted in other different particles as occurs with the mechanism of ALPS-photon conversion \cite{masaki}.
Depending on the evaporation rate or other axion conversion phenomena, the oscillatory regime may be transient and disappear when the DM density decreases if not compensated by the gravitational capture of DM from the star.

If we take in consideration in the stress-energy tensor also the properties of matter present in the star, mostly fermions, one can have a more complete description with a model of mixed fermion-boson star whose stability and oscillatory modes now depend on the central densities of the bosons and fermions there present, evolving from oscillatory to steady dynamical states of the stellar structure and vice versa and including the effect of viscosity damping \cite{valdez}.

In this work we will mainly focus on the axion particle (or ALPs). The other types of DM particles that are not supposed to introduce oscillatory behaviors on the host star obviously will not be discussed here.

\section{Real frequencies and ``ghosts'' introduced by the data analysis}

Borra and Trottier observed ultrafast light modulations in stars \cite{borra} and galaxies \cite{borragal} that interpreted as SETI signals then Tamburini and Licata suggested that these modulations could be signatures of dark matter \cite{tamlicDM} assuming that the frequency $f_S$ is real and not introduced by the data analysis. In the following years other studies were made on a few samples of these sources with mixed results. In the following we will summarize the results present in the literature and discuss them to see under which conditions the DM model can be applied and if all the frequencies of the ultrafast modulations detected so far do have a real physical origin.

To this aim, we now consider for the sake of simplicity a toy model, where a permanent single bosonic aggregation of ALPs with a single particle with mass $m_a$ is piled up in the core of a star with spectral classes from F to K described in an oversimplified way by a polytrope with index $n$ and pressure-density relationship $P \propto \rho^{(n+1)/n}$ \cite{clayton,padma,iben}.

This aggregation of ALPs is expected to behave like a boson star, characterized by a single oscillatory behavior and a single frequency that can give rise to the ultrafast modulation observed by Borra and Trottier. In the oscillating star, breathing behavior is both in the dark matter core and in the spacetime, due to a time dependent stress-energy tensor that gives rise to long-term stable oscillating geometries at a single given frequency that can be in principle experimentally detected. 

If the core of the star does not have convective motions and the energy transfer is radiative like in the stars from the F down to the M spectral classes, the DM core is hosted inside the star without mixing of matter and can start oscillating transferring the oscillations to the spacetime and the hosting star.
In this way, the static stability of the stellar structure is perturbed by the oscillating DM core located in its center \cite{kolb,visinelli17,visinelli21,kumar} oscillating at the characteristic frequency $f_S$ that depends on the mass of the axion $m_a$,
\begin{equation}
f_S= k~2.5 \times 10^{14} ~ \frac{m_a c^2}{eV} ~ Hz.
\label{eq1}
\end{equation}
and $k$ is a positive constant. The spectral fluctuations with frequency $f_S \simeq 0.607~\mathrm{THz}$ (t$_{sec} =1.64 \times 10^{-12}$) revealed in the SDSS spectra by Borra and Trottier would be due to the breathing effects induced by axion-like particles with an upper mass value $m_a \sim 2.4 \times 10^{3}~ \mathrm{\mu eV}$. 
These mass values $m_a$ associated to $f_S$ overlap the mass range of axions from Lattice QCD simulations \cite{qcd,qcd2} when $2 \leq k \leq 100$ and result in agreement with the expected axion detections in Josephson junctions and solar observations \cite{beck,kims} with the SMASH axion model \cite{balles} and with the recent results from laser interferometers \cite{aiello}. This is a known story and may leave us with the doubt that $f_S$ could be a ``fake'' frequency introduced by the data reduction process.

Some hints about the effective physical origin of $f_S$ is discussed taking in account the other two frequencies discovered by Borra and from the re-analysis of their data and data reduction procedure found in the literature and here discussed.
From the analysis of other spectra obtained from the light of SDSS galaxies, Borra found two additional major frequencies in $223$ galaxies over a sample of $900,000$ \cite{borragal}. All the spectral lines and frequencies were redshifted according to the concordance cosmological model \cite{planck}.
Once corrected by the redshift, the two frequencies resulted to be $f_{1,G} \simeq 9.71$ THz, corresponding to $1.03 \times 10^{-13}$ sec and $f_{2,G} \simeq 9.17$ THz, corresponding to $1.09 \times 10^{-13}$ sec. 
Following \cite{tamlicDM}, from Eq. \ref{eq1} if these frequencies correspond to the presence of additional DM fields, then one recovers the following upper mass values $m_{G,2} = 2~\mathrm{meV}$ and $m_{G,1}= 36~\mathrm{meV}$ of two hypothetical ALP fields. These values result to be beyond the upper limit expected from Lattice QCD simulations, even if some different theoretical constructions may admit the extension of the the mass window for the axion QCD model to those values and bejond \cite{diluzio}.

Let's see if and which frequency could have been introduced by the data reduction process. The three frequencies $f_S$, $f_{1,G}$ and $f_{2,G}$ were found by Borra and Trottier in both the SDSS catalogues always using the dubbed Spectral Fourier Transform (SFT) technique, a method developed to detect and characterize the presence of ultrashort pulses.
What is immediately evident is that $f_S = f_{1,G} - f_{2,G}$. This is clearly a frequency-beating relationship. By definition, the beat frequency is equal to the absolute value of the difference in frequency of two waves. Thus, either $f_S = | f_{1,G} - f_{2,G} |$ or $f_{1,G} = | f_{2,G} - f_S |$ or $f_{2,G} = | f_{1,G} - f_S |$. This implies that either $f_{1,G}$ is the beating between the frequency $f_{2,G}$ and $f_S$ or $f_S$ is the beating between the frequencies $f_{1,G}$ and $f_{2,G}$ and so on.
This combination of frequencies and their slow drift and modulation in time is similar to what occurs in acoustics when playing simultaneously two strings in a violin generating the Tartini's third tone \cite{tartini} whose real existence was scientifically demonstrated by Hermann von Helmholtz \cite{vonhelmholtz} and recently discussed for the presence of many other tones with different frequencies that also drift with frequency in time in the violin playing \cite{caselli1,caselli2}.

Now that we have the three frequencies and the beating relationship together with a toy model to play with, as first step, we have to exclude the hypothesis that $f_{S}$, $f_{1,G}$ and $f_{2,G}$ are not introduced by the windowing made with the upper and lower values of the instrumental windows in the SDSS frequency spectrum or caused directly by a combination of them.
The windowing origin of all the three frequencies is excluded as the two spectrographs worked in the frequency ranges $(380 nm <  \lambda  < 615 nm)$ and $(580 nm <  \lambda <  920 nm)$. The limits of the spectral range are defined by the first bin at $N=1$ that corresponds to $2.1538 \times 10^{-15}$ sec viz., $4.6430 \times 10^{14}$ Hz and the latest bin at $N=1950$, which corresponds to $4.1999 \times 10^{-12}$ sec, i.e., $2.3810 \times 10^{11}$ Hz. It is a wide range spanning three orders of magnitude both in frequency and time with no clear correlation with the three frequencies, from the visible violet to the near infrared.

To see, instead, whether the SFT analysis introduced some frequency ghosts and to see whether the three frequencies are not all numerical ghosts, we now analyze the other results present in the literature: Hippke demonstrated that $f_{2,G}$ is a spurious frequency using the same SFT method by Borra and Trottier \cite{hippke}. The key point that demonstrates that $f_{2,G}$ is introduced by the SFT analysis is that this frequency was found again in the galactic spectra as expected but, surprisingly, $f_{2,G}$ was also found by analyzing the synthetic solar spectrum by Kurucz \cite{ku1,ku2}. This ``artificial'' solar spectrum is a solar spectrum artificially built from tabulated values of known and calculated solar spectroscopic lines, with transitions and related phenomena expected to occur in the Sun. An example of the lines, reconstructed with high accuracy, present in the synthetic spectrum are the lines of the Ca II doublet, which are influenced by strong departures from the local thermodynamic equilibrium, three-dimensional radiative transfer and partially coherent resonance scattering of photons in the outer layers of the Sun.
Finding $f_{2,G}$ in the artificial solar spectrum is a smoking-gun proof, as this artificial solar spectrum cannot have neither SETI signals nor DM signatures reported inside.

To explain this first result more in detail, the SFT analysis of the synthetic solar spectrum by Hippke puts in evidence a high peak with power $\mathrm{pow}_{2,G} \simeq 22.4$, corresponding to $f_{2,G}$ ($1.09 \times 10^{-13}$ sec) but no power in correspondence of $f_S$, for which the power is about $\mathrm{pow}_S \simeq 0.6$ on the order of the background noise, embedded in the noise near the deepest region between two peaks, a and b, located at the time intervals and with power (t$_{a} =1.49 \times 10^{-12}$ sec, $\mathrm{pow}_a = 2.21$) and (t$_{b}=1.70 \times 10^{-12}$ sec, $\mathrm{pow}_b =  2.34$). 
The term ``pow'' represents the power of the peak in the Fourier spectrum. Summarizing, only the ``fake'' ghost frequency $f_{2,G}$ is produced by Hippke's analysis of the synthetic solar spectrum but neither $f_{1,G}$ nor $f_S$ are present. 
The SFT analysis by Hippke showed that this frequency corresponds to the deepest point with $\mathrm{pow} < 2.65$ between the peak of $f_{2,G}$ and another one at (t$ =1.02 \times 10^{-13}$ sec, $\mathrm{pow} = 8.07$) in the Fourier spectrum.
This would suggest that at least only one of the two latter frequencies can be real and the other is the beating with the ghost $f_{2,G}$.

Let us make some consideration about the ``fake'' frequency $f_{2,G}$. It  was artificially introduced by the SFT technique with a probability $\sim 0.05\%$ to happen by coincidence, due to the non-random spacings of certain spectral absorption lines that are in common with all the spectra we observe (see the discussion in the appendix for more details about the galactic spectra).
The spectral lines present in both the galactic and synthetic solar spectra form a comb-like structure characterized by a regular set of distances between each spectral line and SFT erroneously identifies this periodicity as a real modulation due to ultrashort pulses in the spectra with frequency $f_{2,G}$. More precisely, $f_{2,G}$ emerges from the comb-like structures of the spectral lines that are common in all the spectra up to now analyzed (stellar spectral classes F to K) such as the Balmer hydrogen series or the one time ionized calcium lines that are also present in the synthetic solar spectrum. The Sun is, in fact, a G2 type star. 
This suggests that stable absorption lines common in all the spectra, including the artificial solar one should be those responsible for the ghost $f_{2,G}$.Any galactic spectrum is the result of the integration of all the spectra of the stars of that galaxy, mainly made of stars with spectral classes from F to K.

Now one has to understand which of the two frequencies is the ``real'' one or if both are fake as well. Let us consider the frequency $f_S$. Certain models of particle physics suggest that the mass range discussed in \cite{tamlicDM} would favor $f_S$ to be interpreted as a real frequency related to DM, but this is not enough as we do not know whether the axion or the ALPs exist and what are their actual physical properties. This frequency has also been revealed in some of the galactic spectra presented in \cite{borragal} but is mainly found through the beating of $f_{1,G}$ with $f_{2,G}$. 
An important clue for the existence of $f_S$ comes from the comparison between the analysis of the synthetic solar spectrum ($f_S$ was not detected there) and the real solar spectrum ($f_S$ was instead revealed). 
Differently from the other two frequencies $f_{1,G}$ and $f_{2,G}$ both detected in the galactic spectra, the probability of finding the peak at $f_S$ in the sample of stars of our galactic halo is much lower, $0.009\%$, versus the much higher probability of having $f_{2,G}$ which is about $\sim 0.05\%$, about six times that can couple with some power that can be found in the galactic spectra in correspondence to $f_S$ giving rise to $f_{1,G}$ with always higher power than $f_S$ because of the always present fake frequency $f_{2,G}$ introduced by the SFT analysis due to the properties and structures of spectral lines present in the galactic spectra. This is the reason why $f_S$ was less evident in the galactic spectra with respect to the other two frequencies: the highest power of $f_{2,G}$ beating with the lower $f_S$, according to the probability of detection (the galactic spectra are the integration of billion of stellar spectra and other sources), of course give rise to a more evident beating in $f_{1,G}$.
A more detailed discussion about the properties of the synthetic, real solar spectra and stellar and galactic spectra is reported in the appendix.

In favor of $f_S$ again: this frequency was observed in the real solar spectrum taken with the solar spectrometer installed at the International Scientific Station of the Jungfraujoch by the University of Li\`ege \cite{curdt,delb,delb2}, together with $f_{1,G}$, by using the dubbed SFT technique.
Peering more in detail in the real solar spectrum, we notice that the SFT analysis of a real solar spectrum clearly shows the presence of a small bump in correspondence to $f_S$ with the following properties: (t$_{S} =1.64 \times 10^{-12}$ sec, $\mathrm{pow}_S = 6.47$). 
This peak is well above the background noise and is present the peak $f_{1,G}$ with (t$_{1,G} =1.03 \times 10^{-13}$ sec, $\mathrm{pow}_{1,G} = 8.04$) too, which results well detectable, and in this way finds another confirmation that it is the beating between the highest peak of the Fourier spectrum, $f_{2,G}$, and $f_S$.
The detection of $f_S$ in the real solar spectrum implies that most likely the synthetic solar spectrum analysis could not take in account all the features of the solar lines including their variations or fast oscillations as those due to DM or to flare or plasma activities. In fact, not all the transient phenomena of the Sun are included there. This means that $f_S$ is found only in real spectra as it is present in all the spectra of the $236$ SDSS main sequence stars with spectral lines similar to our Sun, from F to K spectral types most of them with a stable radiative core that can host a DM core.

Of course, as we cannot a priori exclude the presence of SETI signals in our solar system for obvious reasons, this result favors the idea that $f_S$ may be related to physical processes in the Sun like the DM effects here discussed or can be ascribed to high energy emissions from microflares/flares involving plasma and magnetic fields, introducing a beating frequency from a still not completely known effect of flare activities that are known to exhibit pulsating emission. Similar behaviors have been observed in our Sun with frequencies from $0.5$ to $1$ pulses per second in the sub-THz (up to 405 GHz) and in the 30 THz band with a positive slope in the spectral component at the highest observable frequencies \cite{thz1,thz2,thz3,thz4,thz5}.

\subsection{Frequency drift and  from different observations}

\begin{figure}
    \centering
    \includegraphics[width=1\textwidth]{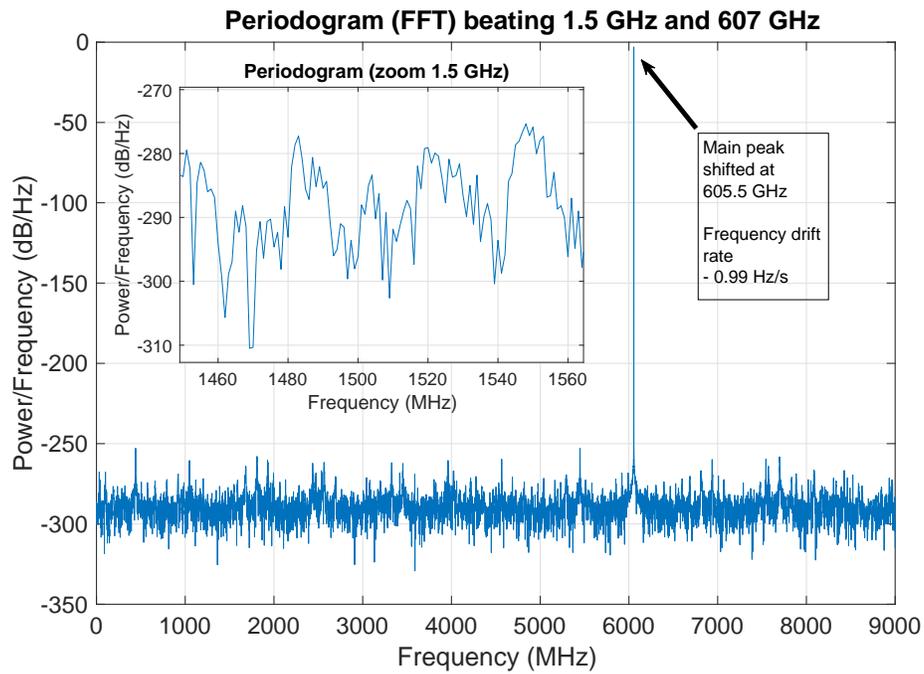}
    \caption{Periodogram of the beating between $f_S$ and a typical flare activity at with varying phase at $f_O=1.5$ GHz (numerical simulation). We obtain the same time drift as observed in the deep-learning SETI search for extraterrestrial technosignatures \cite{ma} from Eq. \ref{battona}.}
\label{fig1}
\end{figure}

Following this line of clues, if the frequency $f_{S} = 0.607~\mathrm{THz}$ is present in the spectrum of these stars, it can be ascribed to one of the causes of the frequency drifting observed in the stars HIP 62207 (10 Canum Venaticorum - HD 110897), a debris disc star \cite{marshall} and HIP 54677 (HD 97233). This frequency drifting was put in evidence by the deep-learning search for techno-signatures in $820$ nearby stars looking for narrowband Doppler drifting of radio signals. These stars represents the actual intersection of the set of stars observed by Borra and Trottier  and to the sample where possible SETI signatures have been found by Xiangyuan Ma and collaborators \cite{ma}. We superimpose $f_S$ with the frequencies there observed $f_O$ in the periodograms. By varying the phase change in time between $f_S$ and $f_O$ in the numerical simulations one obtains frequency drifting of $-0.4998$ up to $-0.9996$. in Fig \ref{fig1} is shown the periodogram of the beating between $f_S$ and a frequency of $1.5$ GHz with a slow phase modulation of $1$ Hz from Eq. \ref{battona}. 
\footnote{A debris disc star is a star with an observable relic of the planetesimal formation process it, which is analogous to the Edgeworth-Kuiper belt of our Solar system, making it a good candidate for hosting exoplanets. }.

\begin{table}[!h]
\begin{center}
\begin{tabular}{|c|c|c|c|c|} 
\hline
 Star & Drift rate & Simulated & Frequency & SpT \\ 
\hline
 HIP 62207 & $- 0.05(10)$ Hz/s & $-0.0458$ & $1351.62$ MHz & G0 V \\ 
\hline
 '' & $- 0.126(10)$ Hz/s & $-0.1260$ Hz/s & $1724.97$ MHz & '' \\
\hline
HIP 54677 & $- 0.11(3)$ Hz/s & $- 0.1136$ & $1372.99$ MHz & K4V \\
\hline
'' & $-0.11(3)$ Hz/s & $- 0.1136$ Hz/s & $1376.99$ MHz & ''  \\
\hline
\end{tabular}
\caption{\label{tab1} Properties of the two halo stars with peculiar frequency drift from \cite{ma} in the deep-learning search for technosignatures of 820 nearby stars for SETI. These two stars also belong to the sample of $236$ SDSS stars interpreted as SETI candidates by Borra and Trottier \cite{borra} or due to dark matter \cite{tamlicDM}, characterized by the presence of the frequency $f_S = 607$ GHz.}
\end{center}
\end{table}

%

The presence of the frequency $f_{S}$ can be the result of the frequency modulation of the observing events at the GHz frequencies and the $6.07$ THz frequency found by Borra and Trottier and interpreted as SETI signals \cite{borra,borragal} or as Axion dark matter signatures by us \cite{tamlicDM} modulating the frequencies $1.3 -- 1.7$ GHz of the observed phenomena causing frequency drifting. As for the multiple Tartini tones in the sound waves from a violin, also in electromagnetic waves the frequency drift of a low-frequency signal can occur due to beatings with another very high-frequency signal. This phenomenon is known as "beat frequency" or "beat phenomenon." When two signals with different frequencies interact with each other, they can produce a new signal with a frequency equal to the difference between the two original frequencies.

In our case, observing a low-frequency signal with frequency $f_{low} \sim 1.5$ GHz together with a very high-frequency signal with frequency $f_S = 0.607$ THz, there will be an impact on the low frequency signal: the high-frequency signal is significantly higher than the low-frequency signal, the beat frequency will be the difference between the two, and when a modulation in phase is present, this causes the low-frequency signal to appear to drift in frequency. This drift occurs because the beat frequency is effectively modulating the low-frequency signal, causing it to appear as if it is changing in frequency over time.

If the phase of the low-frequency signal is variable in time with respect to the higher frequency one, it will introduce additional complexity to the beat phenomenon. The interaction between signals of different frequencies, along with varying phase relationships, can lead to interesting effects in the resulting beat signal, the Frequency Shifting and Drifting: Varying phase relationships can cause the beat frequency to shift and drift over time. The apparent frequency of the beat signal will not remain constant but will change as the phase difference between the two signals changes. a general description of the beat phenomenon when the phase of the low-frequency signal is variable with respect to the higher frequency one.

To describe the beat phenomenon with variable phase, we can set up a toy model with two sinusoidal signals, a high-frequency signal with frequency $f_{high}(t) = f_{high} + \delta f_{high}(t)$, where $f_{high}$ is the nominal frequency, and $\delta f_{high}(t)$  represents small variations in frequency over time and a low-frequency one with $f_{low}(t) = f_{low} + \delta f_{low}(t)$ with modulating phase. 
The resulting beat signal can be described as the product of the two signals, 
\begin{equation}
B(t) = A_{high}(t) A_{low}(t) \cos [2\pi (f_{high}(t) - f_{low}(t))~t + \phi (t)] 
\label{battona}
\end{equation}
where $A_{high}(t)$ and $A_{low}(t)$ are the amplitudes of the high-frequency and low-frequency signals, respectively. The two quantities $f_{high}(t)$ and $f_{low}(t)$ are the time-varying frequencies of the high-frequency and low-frequency signals, respectively. The varying phase difference between the two signals at time $t$ is $\phi (t)$.
The phase difference $\phi (t)$ will determine how the two signals align with each other, and its variation over time can introduce amplitude modulation to the beat signal. The beat signal will have a frequency equal to the absolute difference between the time-varying frequencies: $| f_{high}(t) - f_{low}(t) |$, which can shift and drift over time due to the variations in the two input frequencies.

\subsection{The stability in time of the ultrafast oscillations}

As discussed before, these oscillatory models can be transient because of DM core evaporation or other causes and obtain, under this hypothesis, additional information about the evaporation process or axion conversion that occurs in a very short time interval of about $10-18$ years, indicating that either the DM particles are ultralight and evaporate in a very quick time and, because of this, they are clustered in knots that stars cross during their galactic motion acting like gravitational DM vacuum cleaners and oscillating only for a very short time of tens of years only. If this hypothesis is valid, the conditions to set up these oscillatory motions would result quite difficult and the oscillatory regime of ultrafast oscillations would result unstable and transient with respect to the time life of a star, unless the star is always surrounded by the DM. The presence in the sample of a M5V fully convective star, would suggest the presence of an axion cooling mechanism due to the evaporation of the DM core and/or rapid flare activities. 
As this oscillatory regime appears to be not to last for a long time, assuming valid the DM hypothesis, either the star capturing the DM acting as a ``gravitational vacuum cleaner'' cannot sustain the evaporation DM core evaporation process because the spatial concentration of ALPs is not sufficient for a stable regime or DM has a clumpy distribution in space at the galactic scale \cite{diemand,inoue} and between galaxies in galactic clustering \cite{rogers}, in favor of ALPs particles. As already said, the other alternative hypothesis is that a more realistic core made with a mix of matter (fermions) and DM (bosons) presents a set of transient regimes from stable to oscillatory to a route to chaos depending on the matter/DM mixture \cite{brito2}.  
A precise result can anyway be obtained with a more detailed model of DM trapped core and DM distribution, which goes beyond the purpose of the present work.

Excluding the hypothesis of SETI signals that they cannot be always be detected back as they are not always transmitting and may be doing something else after the first transmissions, this transient regime could be due to the presence of DM or flares/microflare activities as in M5V stars, one present in our sample, that are fully convective and in principle could not host a stable oscillating DM core inside. 

In favor of transient physical phenomena are the results obtained during a SETI campaign, with the Automated Planet Finder and high-resolution optical Levy Spectrometer. Three stars have been observed without finding the same power at $f_S$ as in Borra and Trottier results. While for TYC2041-872-1, a F9V type star no signal is present at all, for TYC2037-1484-1, a G2V star, and TYC3010-1024-1, F9V, even there is no presence of a clear high peak at $6.01$ THz, one notices the presence of a smooth small wide bump with some power just below $1\%$ of the continuum intensity around $f_S$, which is below the sensitivity of the instrument and at the limit to be considered an acceptable result. Similar features can be seen in the simulations obtained by superimposing a sinusoid with frequency $f_S$ to the real spectra \cite{isaacson} obtaining similar results when the power of the ultrafast periodic oscillation there simulated is $\sim 1\%$ and limited by the sensitivity of the instrument.

Following these results, one can conclude that either the oscillations at $f_S \simeq 0.607$ THz are due to (or enhanced by) the effects of the SFT data reduction process, or their non-detection is due to the different data processing procedures and different noise levels.
A simpler explanation could be that these oscillations regimes are transient modes as indicated by the results of the real solar spectra and from the two MS stars, TYC3010-102 4-1 and TYC2037-148 4-1, where a small amount of power was found in their Fourier spectra, but not found in the case of the star TYC2041-872-1. 

Instead, the SFT re-analysis by Hippke of the same spectrum of the F9V star (TYC 2041-872-1) of the SDSS catalogue, which was found initially oscillating by Borra \& Trottier, finds a peak in the periodogram significantly above the noise floor with power at $0.607$ THz, confirming the existence of $f_S$ at that observational epoch.
The star shows some residuals of the presence of the $f_{2,G}$ line and almost a power, $\mathrm{pow} = 5.57$, for $f_S$, which is almost one unit bigger in the Fourier power spectrum than what is detected in the solar spectrum. This suggests that some physical changes have been occurred in the source.

Adopting the DM hypothesis, the presence of a bosonic core in a star could only oscillate for certain periods when the boson core can behave like a boson/fermion mixed star, known to present stable and metastable oscillating regimes in time \cite{valdez} as already discussed. 
In this case, $f_S$ could indicate the presence of a transient behavior different from that of an ideal boson star made with axion DM. The transient behavior observed can be also attributed to an axion core evaporation or to axion-photon coupling and production as hypothesized in massive stars, including the conversion mechanisms of axions with photons such as photon/matter interactions \cite{raffelt}, even if high magnetic fields are required \cite{friedland,millar}. 
Anyway, depending on the properties of the axion, some authors think that even the coupling with the stellar magnetic fields that permeate down to the radiative zone should be able to produce photons reducing the axion core \cite{gnedin}, with a tiny extra production of energy from the star. 
In this case the Sun could be one of the brightest sources and detector of axions from the axion-photon coupling that produces axions with large mass through the inverse Primakoff scattering of thermal photons embedded in the electromagnetic field of the solar plasma \cite{dent} favoring $f_{1,G}$ instead of $f_S$. 
The Majorana Demonstrator experiment sets new limits on the axion-photon coupling, $g_{a~\gamma} < 1.45\times 10^{-9} GeV^{-1}$ with a c.l. of $95\%$ for massive axions with mass in the range $1.4 < m_a < 100~eV/c^2$, whose upper limits go outside from our mass range \cite{axionsun}, finding agreement in its lower values with the mass ranges here found. 

Another cause for the transient behavior of the oscillations could be due to the onset of additional frequencies or instabilities inside the stellar core due to multiple states of the same field or multiple fields \cite{liebling} that can temporarily reduce the power associated to $f_S$ that can be explained by particle theory. This also recalls the behavior of models of fermion-boson stars that can have transitions from stable to unstable configurations and vice versa \cite{valdez} and both amplitude, oscillations and stability of the oscillatory regime depend on the mixture of matter/DM present in the core of the star. 
This model agrees with most of the stars sample which corresponds to the $\sim 1 \%$ MS stars in the halo, population II and metal-poor \cite{jofre} and almost all belong to F-K classes known to have a radiative stable nucleus. In this case oscillatory regimes due to DM can occur. 
In favor of this DM scenario is that F9 is the dominant spectral type in which the distribution is peaked, with mass $M \sim 1.2~M_\odot$ and the energy transfer is almost totally radiative, like in the ideal model described by a polytrope.

Only a campaign of observations in the sub-THz regime of the Sun at the frequencies $f_S$ and $f_{1,G}$ could clarify better this scenario.

\section{Discussion and conclusions}

Following \cite{tamlicDM} we analyzed the common features of the galactic and stellar spectra of the SDSS that presented an oscillatory regime at the frequency of $f_S=0.607$ THz and other two frequencies, $f_{1,G}=9.17$ and $f_{2,G}=9.71$ THz, revealed with the the dubbed Spectral Fourier Transform (SFT) technique, developed to detect and characterize the presence of ultrashort pulses \cite{borra,borra2010,borra2012}. A clear relationship of frequency beating between the three frequencies is evident: either $f_{1,G}$ results to be the beating of $f_S$ with $f_{2,G}$ or $f_S$ is the beating of $f_{1,G}$ with $f_{2,G}$. 

An important clue for the real physical origin of $f_S$ is given by the analysis of the artificial solar spectrum and a real solar spectrum using the same SFT technique of Borra and Trottier.
From the SFT analysis of the Kurucz synthetic solar spectrum, compared with other techniques, it has been shown that the observed peak at $9.71$ THz ($f_{2,G}$) was originated by the SFT procedure \cite{hippke}, caused by the geometry of the spacing of the absorption spectral lines in the SDSS spectra. Noteworthy, the other two frequencies were not detected in the artificial spectrum.
Instead, in the real solar spectrum obtained with the solar spectrometer at the International Scientific Station of the Jungfraujoch the SFT procedure revealed, together with $f_{2,G}$, the presence of the two frequencies $f_{1,G}$ and $f_S$, a clear indication that something was missing in the artificial solar spectrum even if the main spectral lines and, above all, their spacings, are correctly reported.
The spectral lines responsible for the ``fake'' frequency $f_{2,G}$ are the main spectral lines in common with the other main sequence halo stars and galaxies of the SDSS survey. 

One can conclude that, if $f_S$ were real, it must be originated by the same physical phenomenon occurring in these galaxies, stars and in the Sun and this has nothing to deal directly with the spacing of the main spectral lines like those of hydrogen and calcium present in most of all the stellar spectral classes that can introduce numerical ghosts as in the case of $f_{2,G}$.

Other observations with different data reduction techniques showed in the Fourier spectrum the complete lacking of $f_{2,G}$ and the presence of a low peak in correspondence with $f_S$, just below the $1\%$, the value to which corresponds the smallest detectable value due to the sensitivity of the instrument \cite{hippke,isaacson}, suggesting that either SFT technique is not reliable or we are dealing with transient phenomena in the stellar structures such as forming and evaporating axion bosonic cores or a core with a mixture of bosonic and fermionic matter and dark matter of which theory expects transient oscillatory scenarios and temporarily stable structures evolving from static to oscillating and vice versa or to a route to chaos. 
Assuming $f_S$ real, then one finds an agreement with the expected axion masses as already discussed in Ref. \cite{tamlicDM}, including the recent results of the gamma ray burst GRB221009A, for which $m_a = O(0.01 - 0.1)~\mathrm{\mu eV}$ \cite{GRB1,GRB2,GRB3}, for with $k > O(10^3)$.

The discrepancy between the predictions of the Standard Model of particles and the measured the muon g-2 anomaly suggests the presence of a new physics involving the muon or  around or below the weak scale and a better study of the effects of vacuum polarization of the Standard Model. In fact, if the hypothesis of DM induced ultrafast oscillations are valid, the upper mass value found is in contrast with certain models that explain the anomalous magnetic dipole moment of a muon, also known as muon g-2 anomaly \cite{muon1,muon2} involving the axion \cite{muon3}.
In certain scenarios these models would require couplings between the Standard Model leptons and the axion, on the order of $(25 - 100)$ GeV and a heavy axion with mass from the $10$ MeV up to the $10$ GeV scale, which is much beyond of the axion (or ALP) model here assumed. 
The models for the muon g-2 anomaly that consider the loop effects of an axion that couples to leptons and photons would mostly need need large axion masses and couplings \cite{muon4} including massive fermion fields with large axion-lepton couplings.
If these results will be confirmed by future experiments, a new set of ALP particles and fermionic fields are expected to contribute to the DM scenario \cite{muon5}. An ultralight axion mass then would imply slightly new physics to explain this anomaly, like another family of ALPs with extremely high mass values, taking in account all the possible theoretical \cite{muon6} and experimental issues related to this challenging and key measurement for particle physics.

An alternative explanation to the onset of ultrafast oscillations could be due to plasma activities and magnetic fields or transient flare or microflare activity, as expected in our Sun. In favor of that is the presence of an M5V star in the SDSS sample of $236$ stars which is not expected to can host a stable DM core as it is fully convective but that shows ultrafast oscillations, unless a very efficient axion cooling mechanism is present due to the evaporation of the DM core, stabilizing for a short time the inner stellar structure.
both these hypothesis can explain the transience in the oscillatory regime of certain stars when re-observed.
 
To prove the flare/microflare hypothesis we would need further and deeper experimental investigations of the high-energy processes in solar and stellar flares occurring during the impulsive phase of the flares.
New observations in the Sun in the THz and sub-THz bands and more precisely in the solar emission around $0.607~\mathrm{THz}$ may provide better insights in this and in the DM scenario through a detailed spectral analysis in the sub-millimeter wavelength with high time resolution.

Last but not least, if $f_S$ would be caused by SETI activities, the ultrafast oscillations would be of course a transient phenomenon, present in stars with planets in the habitable zone and in our solar system, but as $f_S$ has been observed in the Sun, this may favor the DM hypothesis unless thinking about other speculative scenarios that are more suitable for Sci-Fi writers.

\subsection*{Acknowledgments}
FT thanks Maestro Stefano Zanchetta for having introduced him on how to generate and handle Tartini's tones with a violin A. Guarneri and other instruments.
\\

\section{Appendix}
\subsection{Properties of the spectra of the source samples:} 
Let us discuss the main features present in the spectra where the three frequencies were detected and the stellar spectral classes of the stars considered. 
The stars analyzed by Borra and Trottier and in the other following works \cite{hippke,isaacson,ma} are main sequence (MS) stars for which the energy flux in the inner core is carried out by radiative transfer and the matter there inside behaves like a static fluid. If ALPs pile up in the stellar core \cite{brito,brito2}, then spacetime and the local density of the radiative stellar core start an oscillatory regime with one or more periods that depend from the properties of the scalar field. The oscillatory regime remains stable in time if the main DM core remains stable with enough particle concentration to modify the relativistic stress-energy tensor and onset the ultrafast oscillation regime of matter and spacetime \cite{brito,brito2}. Axion-photon conversion, DM core evaporation and other different interactions between axions and the fermions that makes the matter of the star will modify the stress-energy tensor with the result that the oscillatory regime will be either damped or disappear or experience transitions from static to oscillatory with a possible route to chaos as already discussed in the main text. The only exception is the MV5 star which is expected to be fully convective, either with a DM dominated energy transfer or suggesting that these ultrafast oscillations can be originated by flare activities, that may occur in this spectral class.

\textbf{Synthetic solar spectrum}. The synthetic solar spectrum is artificially built with data from spectral libraries that provide lists of the atomic and molecular lines and other information needed to model the spectrum of the Sun at best. As it is known, synthetic spectra still do not reproduce all the features of the observed lines; several spectral transitions and some minor lines may be missing and the profile of certain lines of the computed synthetic spectra can have, in the worst case, errors up to the $200 \%$ \cite{kitamura} but not in their positions in the spectrum. In any case most of the fundamental lines such as hydrogen and calcium lines in the solar spectra \cite{scott}, also observed in the galactic spectra are present. Here no ultrafast oscillations can be present and were not detected.

\textbf{Galaxies:} the galactic spectra are the result of an integration of all the stellar spectra and emission mechanisms that can occur in a galaxy; all the stellar spectral classes, nebulae and other sources contribute to the profile of a galaxy's spectrum in proportion with their population. 
The spectrum of a galaxy is at all effects the convolution of the spectra of the stars, each of which can be seen as a tiny signal overwhelmed by noise of the light of all the other sources there present but, by adding many stellar spectra together, the main spectral lines will unavoidably emerge and build up the ``spectral comb'' of spectral lines for $f_{2,G}$ and sum up also the other effects that generate the other two frequencies $f_S$ and $f_{1,G}$.
This comb-correlation effect in the spectral lines remains valid also when the frequencies are redshifted: the redshift effect of the galactic spectral lines is a rigid translation, with a probability $0.025\%$ ($223$ on a sample of $0.9$ million galaxies) as some of the lines that can be held responsible for the presence of $f_{2,G}$ can be recovered from the noise by summing up many spectra of the stars in each of the galaxies. Because of this, the argumentation used to discard $f_{2,G}$ does not apply to $f_S$ and $f_{1,G}$ is nothing but the beating with the spurious frequency $f_{2,G}$. 
Of course we can conclude that when $f_{1,G}$ and $f_S$ are detected in the galactic spectra, differently from the synthetic solar spectrum, the galactic spectra are real spectra and $f_S$ can be supposed to be a real frequency.

Interestingly, the integrated spectrum of the other galaxies presented similar features but the bump at $f_S=0.607$ THz was not always reported in the Fourier galactic spectra even if put in evidence by $f_{1,G}$ that inherits most of the power from $f_{2,G}$, a spurious frequency caused by the data analysis procedure.
The most common spectral lines that are found in the observed integrated galactic spectra are absorption and emission lines that present a smooth progression with the Hubble type classification of galaxies.
The integrated galactic absorption and emission line spectra present a smooth progression with the Hubble type classification of galaxies. Most spiral and irregular galaxies exhibit detectable emission lines of hydrogen, H$_\alpha$, the strongest emission line generated in ionization-bounded HII regions and by stars accompanied by the nebular lines like the doublet [OII], related to the internal kinematics of the ionized gas and found also in B0 spectral type stars. Other nebular multiplets like [NII] and [SII] can be found. These lines are followed by H$_\beta$ with [OIII] \cite{kenni}. Nebular multiplets are not so frequent in F to K stellar spectral classes.

From this, one can find a whole set of common frequencies in all the data analyzed with SFT, from the artificial solar spectrum to the SDSS stars and galaxies. Each single star has its own spectrum with its peculiar lines that depend on the spectral class. Mostly the signals arise in many luminous celestial bodies   having spectra with high S/N ratio in which absorption lines are present.
To form the spurious frequency $f_{2,G}$, these spectral lines has to form a comb-like structure that give rise to a set of distances of absorption lines that result correlated. Amongst these lines, which can be recovered from the noise by summing up many spectra, there are the Balmer H (mainly H$_\alpha$ and H$_\beta$) and Ca II lines, with CaII H \& K and the CaII IR triplet that have in common with the MS stars and the solar spectrum (artificial and real).

\subsection{Common features of spectral classes and spectral lines}

We now analyze the main characteristics of the stellar spectra per spectral class and of the galactic spectra in which a signal was detected to give evidence to the similarities in the spectra and in the distribution of the spectral lines as stars of different stellar class and with different metallicity do exhibit a set of different absorption lines that generates $f_{2,G}$ as spurious line. 
The spectra of those stars, even if different, should have common features that form stable comb structures of spectral lines that generate not only $f_{2,G}$: the power of the corresponding pulse spacings in the Fourier periodogram will vary as well for any different set of spectra but should preserve some power in certain frequencies associated to artificial frequencies like $f_{2,G}$.

In the sample of the $236$ stellar spectra of our galaxy the $f_S=0.607$ THz oscillation was found mainly in F and G stars in a range of spectral classes from A to K and one sample with a M5V star. 
This sample presents a first selection effect due to the SEGUE survey which targeted stars mostly of the F to K-type, cutting off the O and B classes and lowering the percentage of the A spectral class, where hydrogen lines are very strong and dark with a maximum strength for A0, being H lines strong in the temperature range from $4000~^\mathrm{o}\mathrm{K}$ to $12000~^\mathrm{o}\mathrm{K}$. The other relevant lines are those of one time ionized Calcium, Ca II, which start becoming more evident in the late subclasses of F stars to M7 stars, being very weak in A stars. Other ionized metals such as Mg II and Fe II are present.

These MS stars have masses in the interval $0.3 - 1.2~M_\odot$, spectral classes from F to M and are in the halo, which means that they are mainly of population II, metal-poor with low metallicity less affecting the spectral lines than in our Sun which is of population I.
When the stellar mass $M \simeq 1.2-1.3 ~M_\odot$, stars are almost totally radiative and for smaller values of mass, the stellar core starts being surrounded a by a convective envelope, like our Sun. In these stars ideally a bosonic core could form and start its oscillatory motion. This may favor the hypothesis of DM oscillations. 
Instead, MS stars like the M5V star in our sample  ($M<0.3~M_\odot$) are totally convective and with flare/microflare activities. This, instead, may favor the hypothesis that the oscillations could be due to flare activities or dominated by axion cooling. In addition to the M5V star, there are five of them that have spectral class A, supposed to have a convective nucleus. In principle, these types of stars do not fit with the stable oscillating bosonic core model, unless astrophysical effects were temporarily stabilizing the structure of the inner core. 
Some of the mechanisms that can be invoked to stabilize the inner core are stellar rotation, DM influencing the stellar evolution processes such as the energy transport in the inner core \cite{jofre,clayton,padma,iben}: the energy generated by the nuclear reactions is transported away in a very efficient manner \cite{kolbe}. These non-standard axion cooling mechanisms may explain the discrepancies between the observational data and stellar evolution \cite{kolbe,giannotti}, with the result that
their lifetime is shortened and the oscillation phenomena may be transient.
These exceptions might be caused by other astrophysical effects or even by the presence of DM fields that may alter their structure.

In this sample there are no giants, collapsed objects or stars away the main sequence. The selection of MS stars from the other types of stars in the same spectral class occurs via their spectra. MS stars are mainly characterized by the ratios of the magnitudes of the H spectral lines at $4045$ \AA~(H$_\delta$) and $4226$ \AA~(H$_\gamma$). The most sensitive criterion of absolute magnitude is at all effects the Spectrum--Luminosity Indicator in late-type stars with the ratio between the Sr lines at $4077$ \AA~and the luminosity at $4226$ \AA~of Ca I that result different for normal giants and dwarfs. 
For supergiants and giants, instead, one uses the ratios $4077$ \AA~with that of H$_\delta$ and $4171-4173$  \AA~with $4226$ \AA~allowing a very accurate luminosity classification \cite{fernie}. 

Excluding the only one A-class star present in the sample, which is supposed to have a convective nucleus (see \cite{tamlicDM} for more details), the main spectral classes analyzed by the SDSS where these frequencies were found, independently from the metal content and stellar population are the following:
\\
\textbf{F stars:} what we notice in Tab. \ref{tab2} is that the SEGUE selection effect is not ruling the percentage per spectral class of the stars with $f_S$, the dominant spectral class with that signal is the late sublcasses of F type stars ($61\%$), F8-F9 of solar type (F8V and F9V), where there is the limit to the onset of turbulence in their inner cores that may be mitigated by the axion cooling.
In this spectral class hydrogen lines (mainly the series from Lyman H$_\alpha$ to H$_\delta$) are dominant and are present other ionized metal lines that start becoming evident.
\\
\textbf{G and K stars:} just after F9 stars one finds the G-type stars like our Sun (G2V), in which the signal was present, and K-class. Both the classes have similar percentages of stars with $f_S$ present, (17,4 \% and 14.8\%), but their respective population ratio in the Milky Way is quite different different, as the ratio between the number of G stars with respect to K stars is $2/3$.
We can deduce that the percentages of stars with $f_S$ in these two populations is not due by instrumental selection effects or by their populations, indicating that the oscillation at $0.607$ THz needs for additional explanations.
In G stars and, in particular in the subclass G2V, are still present the H and CaII lines together with other absorption lines of neutral metallic atoms and ions that grow in strength as one moves in the lower spectral classes.
\\
\textbf{M5V, wholly convective, with microflare activity.} 
In the M stars the spectrum is dominated by molecular bands, especially of titanium oxide are present together with those of neutral atoms and Ca II, much stronger if the star is active, with flares and microflare activity. 
Differently from these three spectral classes, the only M5V star (id. 2MASS J12160751+3003106 -- Low-mass Star from the Two Micron All Sky Survey \cite{2mass}) in the sample that presents $f_S$ is a MS red dwarf. M5 star is supposed to be totally convective and an oscillating boson core would result perturbed and quite unstable, unless the stellar structure is modified by the presence of axions with a resulting cooler stellar structure.
Microflare activity is present in these stars and it would be a possible candidate for the line $f_S$, even if there is still lacking of observations in the sub-THz band at $600$ GHz also in the Sun and solar flares.
The active fraction of stars of the M class peaks at spectral type M8, where $73\%$ of stars are active and one cannot tell the active stars from non active ones by using the color index only. The activity is revealed by their optical spectra that present interesting features in the H$_\alpha$ emission line, CaII and the higher-energy hydrogen Balmer lines from the chromosphere. With its g-r $= 1.826 \pm 0.064$ index the star is redder than a typical M5V star, for which g-r $= 1.52 \pm 0.13$, indicating that the star may be active and presents Ca II lines as in the other spectral classes. 
More precisely the activity is measured with the ratio of the luminosity emitted in H$_\alpha$ to the bolometric luminosity, $L_{\mathrm{H}_\alpha}/L_{\mathrm{Bol}}$. 
The decline in mean activity strength begins at spectral types M5-M6  \cite{sloan,sloan2}.
Even if these differences are not statistically significant, for an active star one finds a slightly bluer u-g color index and the g-r color index slightly redder when compared to inactive stars. This does not apply for the star 2MASS J12160751+3003106 we are considering. The star is at $168.53$ pc and the fraction of M5 active star of the SDSS sample is quite high, $0.55$, with a temperature T$_{\mathrm{eff}} = 3228 \pm 166~^\mathrm{o}\mathrm{K}$. If the DM hypothesis is correct, the possible explanation is the axion cooling mechanism.

\textbf{Common spectral lines: H and Ca II lines}
In the three classes, F, G, K, the most dominant spectral absorption lines are those of Hydrogen, with weak to very low effects in the spectrum with respect to the earlier spectral classes. Other relevant spectral lines are those of one time ionized calcium, Ca II. In the solar spectrum the two strongest lines are the resonance doublet of Ca II, the H line at $3968.469$ \AA~and the K line at 3933.663 \AA. Ca II presents also a triplet of three strong lines at $8498.018$\AA, $8542.089$\AA~and $8662.140$ \AA~formed in the chromosphere \cite{refId0} in the infrared region.
There are some similarities between the F and G classes because of the presence of ionized metals, whilst G and K spectral classes have in common the presence of absorption lines due to neutral atoms with an effect of presence of spectral lines varying from weak (G) to strong (K). 

\begin{table}
\begin{center}
\begin{tabular}{|c|c|c|c|c|}
\hline
Spectral & Temperature & Absorption  &\% stars & \% w/signal\\ 
 Class & $10^3~^\mathrm{o}$K & lines&  & 236 \\
\hline
A & 7.5 - 10 & H(S), Ca II, Mg II, Fe II & 0.6\% & 1.7 \% \\
\hline
F& 6 - 7.5 & H(w), Ca II &  3\% & 61.9\% \\ 
 & &  ionized M & & \\
 \hline
G& 5.2 - 6 & H(w), Ca II& 7.6\% & 17.4\% \\ 
 & & ionized + neutral M& & \\
 \hline
K& 3.7 - 5.2 & H(W), Ca II (S) & 12.1\% & 14.8\% \\
 & & neutral M (s) & & \\
 \hline
M5& 2.4 - 3-7 & Neutral atoms (s), TiO & 76\% & 0.4\% \\ 
 \hline
\end{tabular}
\end{center}
\caption{Spectral classes, temperature and spectral lines, (S) = strongest, (s) = strong, (W) = weak, (w) = weaker, (M) = metals, i.e., chemical elements heavier than hydrogen and helium (data from Ref. \cite{temp,pop,lines}). At $0.607$ THz there is not a clear correlation between the percentage of stars in our galaxy and that of the detected signal per spectral class, indicating that the selection effect of SEGUE survey is not dominant.}
\label{tab2}
\end{table}



\end{document}